\documentclass[12pt]{iopart}

\usepackage[dvips]{graphicx}
\usepackage{mathtext}       % before inputenc & babel !!!
\usepackage[english]{babel}
\usepackage{citehack}       % after inputenc & babel !!!

\newcommand{\vt}{\textmd{Ti-6.5Al-2.5Mo-1.5Cr-0.5Fe-0.3Si}}
\newcommand{\lb}{\left(}
\newcommand{\rb}{\right)}
\newcommand{\pd}{\partial}

\begin{document}

\title[Liquid pool of VT3-1 titanium alloy during VAR]{Simulation of the liquid pool
for VT3-1 titanium alloy during vacuum arc remelting process.}

% \title[Mathematical Modelling of VAR for Titanium Alloys]
% {Mathematical Modelling of Vacuum Arc Remelting Process for Titanium Alloys.}

\author{E.N.~Kondrashov\dag,\
M.I.~Musatov\ddag,\
A.Yu.~Maximov\dag, A.E.~Goncharov\dag, and L.V.~Konovalov\dag 
}

\address{\dag\ JSC VSMPO-AVISMA Corporation, Parkovaya str. 1, Verkhnyaya Salda, 624760, Sverdlovsk
region, Russian Federation.}

\address{\ddag\ JSC VILS,  Gorbunova str. 2, Moscow, 121596, Russian Federation.}

\ead{evgeniy.kondrashov@vsmpo.ru}

\begin{abstract}

This article describes a simple heat model of the vacuum arc remelting (VAR) process that includes
solution of the nonlinear heat conductivity equation with the nonlinear boundary conditions
which are typical for VAR process. The finite-difference analogue
of the model equations was obtained through the finite volume method. To check the
efficiency of the simplified model that does not include
magnetohydrodynamic phenomena in the liquid metal pool, the comparison has been made of the
numerical calculation of the metal pool depth when
melting the Russian titanium alloy VT3-1 with the results of radiographical tests.
It was established that the model adequately describes the test data
for various melting modes (ingot diameter and current strength).

\end{abstract}


% \submitto{\MSMSE}

% \maketitle

%%%%%%%%%%%%%%%%%
\section{Introduction.}

Today a lot of attention is given to automation of VAR furnaces and the focus is on development of the arc clearance control
flowcharts, feedback controllers, program management systems for furnace electric modes \cite{van}. Together with the aforementioned
issues, the development of the VAR automatic control systems requires a reliable theoretical description of the ingot formation
process that underlies the automatic control system algorithms. The significance of through and reliable theoretical models
of the ingot process formation is determined by the fact that the immediate control of the ingot parameters during solidification
of titanium alloy (in particular, the depth of the metal pool, width of the mushy zone, etc.) using the measuring instruments,
appears to be a problem so far. Development of the theoretical ideas about formation of titanium alloy ingots is rather important
today since there are cases when different ingot-related defects are observed on billets, bars and eventually on critical parts.
Such defects include $\beta$-flecks \cite{mitchell}, zonal segregation \cite{plavka}, tree-like structure \cite{rosenberg} and
dark/light spots \cite{ken} that may be observed in the transverse
section under the macro examination of mill products in different titanium alloys. Very often such defects are related to the
final remelting stage therefore the theoretical description of the processes that take place during solidification of the VAR
melted ingots, as well as understanding of the impact of various melting parameters on ingot metallurgical quality, is undoubtedly
the paramount objective for creation of the fully automated VAR control system.

Many researchers studied the heat processes in ingots during
VAR \cite{jardy-1,jardy-2,bertram}, electroslag remelt \cite{bertram,esr} and continuous casting \cite{malevich}. Today, in view
of development of advanced computers it is getting important to write science-based VAR simulation programs for their use
in industrial environment in order to develop new remelting modes for complex titanium alloys.

Taking into account aforementioned facts we reviewed in this article the mathematical model of the
vacuum arc remelting and compared the theoretical calculation of the metal pool depth with the
radiographical test results obtained for Russian titanium
alloy VT3-1 (\vt) \cite{radiography}.  Taking into consideration the non-trivial complexity
and essential non-linearity of the task, it should be noted that firstly there is no exact analytical solution of the task
and secondly today there are a lot of different methods of approximation of the boundary conditions and building of the difference schemes
for the heat conductivity equation. It is obvious that in such situation the experimental check of the computational solutions
obtained through different methods of approximation is getting important.  Therefore, in this article
we reviewed both aspects of the problem: making the mathematical model and comparison of this
model with the experimental results.

%%%%%%%%%%%%%%%%%%%%%%%%%%%%%%%%%%%%%
\section{Vacuum arc remelting process.}

We will describe the heat flows during the vacuum arc remelting through the heat conductivity
equation that includes release of latent heat within the interval of the alloy solidification.
In order not to involve the complex interface conditions on the phase boundary, we reviewed
the so-called method of apparent heat capacity \cite{malevich} that allows to include
crystallization rate (i.e. heat source) into the system heat capacity as a complementary
additive term \cite{esr}. Such problem statement allows to make calculation in one domain
without ''distinguishing'' the liquid, solid and mushy zone.

Also, from this point on we considered that density, heat capacity, heat conductivity
and heat-transfer coefficient depend on the temperature in an arbitrary way \cite{marchuk}.

Heat transfer equation for VAR ingot can be written in a form

\begin{equation}
\rho(T) C(T) \frac{\pd T}{\pd \tau} =
\frac{1}{r} \frac{\pd}{\pd r} \lb r \lambda(T) \frac{\pd T}{\pd r} \rb +
\frac{\pd}{\pd z} \lb \lambda(T) \frac{\pd T}{\pd z} \rb,
\label{heat-eq}
\end{equation}
where $T$ is temperature, $\tau$ is time, $C(T)$ is heat capacity, $\lambda(T)$ is heat conductivity,
$\rho(T)$ is density, $r$ and $z$ are radial and axial coordinates.

\begin{figure}
\begin{center}
\includegraphics[width=9.0cm]{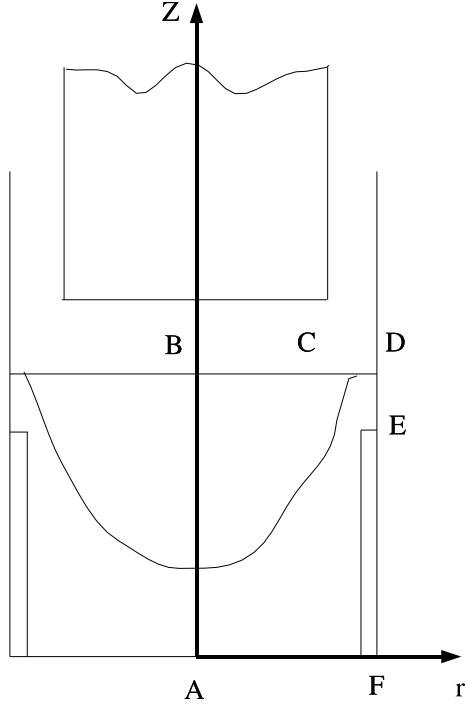}
\end{center}
\caption{\label{var-process} Vacuum arc remelting process.}
\end{figure}

The boundary may be divided into several areas shown in Figure~\ref{var-process}. Such
division is based on physical processes that take place on the ingot surface under VAR.
We will consider those processes and the relevant boundary conditions.

\paragraph{Boundary $AB$.}
On the ingot axis we have the simple symmetry condition:
\begin{equation}
\frac{\pd T}{\pd r} = 0.
\label{cont-bc-AB}
\end{equation}

\paragraph{Boundary $BC$.}
The border $BC $ corresponds to a bath mirror surface which are being under an
end face of the consumable electrode. On this border we shall set temperature.
In the elementary kind the temperature of a surface of a bath is defined by an overheat
above alloy liquidus temperature \cite{jardy-1}:

\begin{equation}
T = T_{L} + \Delta T,
\label{cont-bc-BC}
\end{equation}
where $T_{L}$ is liquidus temperature, $\Delta T$ is overheat.

\paragraph{Boundary $CD$.}
The boundary $CD $ corresponds to a ring gap. Here we also set a boundary condition of the I-st kind

\begin{equation}
T = T_{L} + \frac{D_{cr} - 2 r}{D_{cr} - D_{el}} \cdot \Delta T,
\label{cont-bc-CD}
\end{equation}
where $D_{cr}$ is crucible diameter, $D_{el}$ is electrod diameter.

\paragraph{Boundary $DE$.}
The given boundary corresponds to a zone of a contact belt for which experimental values of a
specific thermal flux $q$ \cite {tet-64} are known, therefore we write down a boundary
condition of II-nd kind

\begin{equation}
-\lambda(T)\frac{\pd T}{\pd r} = q.
\label{cont-bc-DE}
\end{equation}

\paragraph{Boundary $EF$.}
The site $EF$ corresponds to a zone of forming ingot on which has already partially or completely occured
separation of solid ingot from a crucible surface due to passage of process volumetric shrinkage.

Thus three possible mechanisms of heat removal can be observed: 

\begin{itemize}
\item Heat removal by radiation through the formed gap.
\item Heat removal by contact way in places of contact of a surface of an ingot with an internal
wall of copper crucible.
\item The heat transfer by convection through a gas phase in a gap.
\end{itemize}
Three specified mechanisms, it is possible to write down the boundary condition
considering all in the form of 

\begin{eqnarray}
\fl -\lambda(T)\frac{\pd T}{\pd r} = \eta \lb \epsilon \sigma_{0} \lb T^{4} - T^{4}_{cr} \rb +
\frac{k}{d} \lb T - T_{cr} \rb \rb + (1 - \eta) \alpha \lb T - T_{cr} \rb,
\label{1-EF}
\end{eqnarray}
where $\eta$ -- factor describing a share of contact of a surface of an ingot with a crucible, $\epsilon$
-- emissivity, $T_{cr}$ is temperature of inner surface of the crucible, $\alpha$ is
heat transfer coefficient in the contact zone, $k$ is heat conductivity of gas in gap, $d$ is gap. The parameters $\eta$, $\epsilon$,
$T_{cr}$, $\alpha$, $k$ Й $d$ are depending on temperature. For simplification of a record we shall
indicate a following designation

\begin{equation}
\fl \beta_{cr}(T) =
\eta_{cr} \lb \epsilon \sigma_{0} \frac{T^{4} - T^{4}_{cr}}{T - T_{cr}} + \frac{k}{d} \rb + (1 - \eta_{cr}) \alpha_{cr},
\end{equation}
and (\ref{1-EF}) is rewritted in a more compact form
\begin{equation}
-\lambda(T)\frac{\pd T}{\pd r} =
\beta_{cr}(T) \lb T - T_{cr} \rb.
\label{cont-bc-EF}
\end{equation}
Thus, complex process of the heat transfer on border $EF$ is described with the
{\it nonlinear} law (\ref{cont-bc-EF}).

\paragraph{Boundary $FA$.}
On a site $FA $ boundary condition is similar to (\ref {cont-bc-EF}), therefore we shall write down: 

\begin{equation}
-\lambda(T)\frac{\pd T}{\pd r} =
\beta_{bot}(T) \lb T - T_{bot} \rb,
\label{cont-bc-FA}
\end{equation}
where

\begin{equation}
\fl \beta_{bot}(T) =
\eta_{bot} \lb \epsilon \sigma_{0} \frac{T^{4} - T^{4}_{bot}}{T - T_{bot}} + \frac{k}{d} \rb + (1 - \eta_{bot}) \alpha_{bot}.
\end{equation}

Besides it is necessary for us to set the initial condition for the equation (\ref {heat-eq}).
During vacuum arc remelting melt an ingot occurs, i.e. the border $BCD $ moves upwards with
the speed defined by remelting conditions. Melting rate of an ingot is defined by the mass speed
of fusion depending on force of a current, an electrode and a filled crystallizer diameters,
and also on some other parameters.
We shall accept in the beginning of remelting
is already available ingot of some small height with homogeneous distribution of temperature on all volume
is considered to be as the initial condition. 
Initial temperature it is accepted equal $T_{L} + \Delta T$.
For the subsequent moments of time the initial condition is defined by distribution of temperature
during the previous moment of time with additional small layer of an ingot which temperature
is $T _ {L} + \Delta T$.

In the following sections we shall consider discretization of the given problem on a rectangular grid,
including linearization of a system, and decision method of the received the linear equations system.

%%%%%%%%%%%%%%%%%%%%%%%%%%%%%%%%%%%%%%%%%%%%%%%%%
\section{Discretization of heat conduction equation.}

We will consider the most common case when $\lambda(T)$ and $\rho(T)C(T)$ are piecewise continuous
functions. In this case it is convenient to use the integral identity method to the
finite-difference scheme construct \cite{marchuk}. We introduce the heat fluxes along $r$ and
$z$ directions:

\begin{equation}
Q(r, z) = \lambda(r, z)\frac{\partial T}{\partial r},
\label{flux_Q}
\end{equation}

\begin{equation}
P(r, z) = \lambda(r, z)\frac{\partial T}{\partial z}.
\label{flux_P}
\end{equation}
Taking into account these notations, Eq. (\ref{heat-eq}) is rewrited to

\begin{equation}
\rho(T) C(T) \frac{\partial T}{\partial \tau} =
\frac{1}{r} \frac{\partial}{\partial r} \left( r Q(r, z)  \right) +
\frac{\partial}{\partial z} \left( P(r, z)  \right).
\label{main_eq_flux_PQ}
\end{equation}
Integrating the Eq. (\ref{main_eq_flux_PQ}) over the cylindric volume for
$r \in [r_{i-1/2}, r_{i+1/2}]$ and $z \in [z_{j-1/2}, z_{j+1/2}]$
(see Figure~\ref{boundaries}) we can write heat balance equation can be put as follows

\begin{eqnarray}
\fl \int\limits^{r_{i-1/2}}_{r_{i+1/2}} r dr
\int\limits^{z_{j-1/2}}_{z_{j+1/2}} dz
\rho(r, z) C(r, z) \frac{\partial T(r, z)}{\partial \tau}
\nonumber \\
= \int\limits_{z_{j-1/2}}^{z_{j+1/2}} \left[ r_{i+1/2} Q(r_{i+1/2}, z) - 
r_{i-1/2} Q(r_{i-1/2}, z)  \right] dz
\nonumber \\
+ \int\limits_{r_{i-1/2}}^{r_{i+1/2}} \left[ P(r, z_{j+1/2}) - P(r, z_{j-1/2})  \right] r dr.
\label{eq_int_main}
\end{eqnarray}
This equation is {\it exact}. Now we need to evaluate the integrals, which contains
Eq.~(\ref{eq_int_main}). Because we have (\ref{flux_Q}) than

\begin{equation}
\int\limits_{r_{i}}^{r_{i+1}} \frac{Q(r,z)}{\lambda(r,z)} dr = T(r_{i+1}) - T(r_{i}).
\end{equation}
On the other hand

\begin{equation}
\int\limits_{r_{i}}^{r_{i+1}} \frac{Q(r,z)}{\lambda(r,z)} dr \approx Q(r_{i+1/2},z) \,
\int\limits_{r_{i}}^{r_{i+1}} \frac{dr}{\lambda(r,z)}.
\end{equation}
Thus we can write following expressions within the fluxes for internal region of ingot

\begin{equation}
Q(r_{i+1/2},z) \approx \frac{T(r_{i+1},z) -
T(r_{i},z)}{\int\limits_{r_{i}}^{r_{i+1}} \frac{dr}{\lambda(r,z)}},
\end{equation}

\begin{equation}
Q(r_{i-1/2},z) \approx \frac{T(r_{i},z) -
T(r_{i-1},z)}{\int\limits_{r_{i-1}}^{r_{i}} \frac{dr}{\lambda(r,z)}},
\end{equation}

\begin{equation}
P(r,z_{j+1/2}) \approx \frac{T(r,z_{j+1}) -
T(r,z_{j})}{\int\limits_{z_{j}}^{z_{j+1}} \frac{dr}{\lambda(r,z)}},
\end{equation}

\begin{equation}
P(r,z_{j-1/2}) \approx \frac{T(r,z_{j}) -
T(r,z_{j-1})}{\int\limits_{z_{j-1}}^{z_{j}} \frac{dr}{\lambda(r,z)}}.
\end{equation}
Let these expressions be substituted in Eq. (\ref{eq_int_main}) then

\begin{eqnarray}
\fl \int\limits^{r_{i-1/2}}_{r_{i+1/2}} r dr
\int\limits^{z_{j-1/2}}_{z_{j+1/2}} dz
\rho(r, z) C(r, z) \frac{\partial T(r, z)}{\partial \tau}
\nonumber \\
= r_{i+1/2} \left[ T_{i+1,j} -T_{i,j} \right]
\int\limits_{z_{j-1/2}}^{z_{j+1/2}}\frac{dz}{\int\limits_{r_{i}}^{r_{i+1}}\frac{dr}{\lambda(r,z)}} -
r_{i-1/2} \left[ T_{i,j} -T_{i-1,j} \right]
\int\limits_{z_{j-1/2}}^{z_{j+1/2}}\frac{dz}{\int\limits_{r_{i-1}}^{r_{i}}\frac{dr}{\lambda(r,z)}}
\nonumber \\
+ \left[ T_{i,j+1} - T_{i,j} \right]
\int\limits_{r_{i-1/2}}^{r_{i+1/2}}\frac{rdr}{\int\limits_{z_{j}}^{z_{j+1}}\frac{dz}{\lambda(r,z)}} +
\left[ T_{i,j} - T_{i,j-1} \right]
\int\limits_{r_{i-1/2}}^{r_{i+1/2}}\frac{rdr}{\int\limits_{z_{j-1}}^{z_{j}}\frac{dz}{\lambda(r,z)}}.
\end{eqnarray}
Making use the simplest approximation for the integral in the left side of equation we obtain

\begin{eqnarray}
\fl \left( z_{j-1/2} - z_{j+1/2}  \right) \frac{r^{2}_{i+1/2} - r^{2}_{i-1/2}}{2}
\rho_{i,j} C_{i,j} \frac{\hat{T}_{i,j} - T_{i,j}}{\tau}
\nonumber \\
= r_{i+1/2} \left[ T_{i+1,j} -T_{i,j} \right]
\int\limits_{z_{j-1/2}}^{z_{j+1/2}}\frac{dz}{\int\limits_{r_{i}}^{r_{i+1}}\frac{dr}{\lambda(r,z)}} -
r_{i-1/2} \left[ T_{i,j} -T_{i-1,j} \right]
\int\limits_{z_{j-1/2}}^{z_{j+1/2}}\frac{dz}{\int\limits_{r_{i-1}}^{r_{i}}\frac{dr}{\lambda(r,z)}}
\nonumber \\
+ \left[ T_{i,j+1} - T_{i,j} \right]
\int\limits_{r_{i-1/2}}^{r_{i+1/2}}\frac{rdr}{\int\limits_{z_{j}}^{z_{j+1}}\frac{dz}{\lambda(r,z)}} +
\left[ T_{i,j} - T_{i,j-1} \right]
\int\limits_{r_{i-1/2}}^{r_{i+1/2}}\frac{rdr}{\int\limits_{z_{j-1}}^{z_{j}}\frac{dz}{\lambda(r,z)}}.
\label{eq_int_main_2}
\end{eqnarray}
Moreover the integrals containing heat conductivity is simplified up to:

$$
\int\limits_{z_{j-1/2}}^{z_{j+1/2}}\frac{dz}{\int\limits_{r_{i}}^{r_{i+1}}\frac{dr}{\lambda(r,z)}} \approx
\frac{z_{j+1/2} - z_{j-1/2}}{r_{i+1} - r_{i}} \cdot \lambda_{i+1/2,j},
$$

$$
\int\limits_{z_{j-1/2}}^{z_{j+1/2}}\frac{dz}{\int\limits_{r_{i-1}}^{r_{i}}\frac{dr}{\lambda(r,z)}} \approx
\frac{z_{j+1/2} - z_{j-1/2}}{r_{i} - r_{i-1}} \cdot \lambda_{i-1/2,j},
$$

$$
\int\limits_{r_{i-1/2}}^{r_{i+1/2}}\frac{rdr}{\int\limits_{z_{j}}^{z_{j+1}}\frac{dz}{\lambda(r,z)}}\approx
\frac{r^{2}_{i+1/2} - r^{2}_{i-1/2}}{2 (z_{j+1} - z_{j})} \cdot \lambda_{i,j+1/2},
$$

$$
\int\limits_{r_{i-1/2}}^{r_{i+1/2}}\frac{rdr}{\int\limits_{z_{j-1}}^{z_{j}}\frac{dz}{\lambda(r,z)}} \approx
\frac{r^{2}_{i+1/2} - r^{2}_{i-1/2}}{2 (z_{j} - z_{j-1})} \cdot \lambda_{i,j-1/2}. 
$$
After that the Eq. (\ref{eq_int_main_2}) can be rewritten as

\begin{eqnarray}
\fl \rho_{i,j} C_{i,j} \frac{\hat{T}_{i,j} - T_{i,j}}{\tau} =
\nonumber \\
\left( T_{i+1,j} - T_{i,j}  \right) \cdot
\lambda_{i+1/2,j} \cdot
\frac{r_{i+1/2}}{r_{i+1} - r_{i}} \cdot \frac{2}{r^{2}_{i+1/2} - r^{2}_{i-1/2}}
\nonumber \\
-
\left( T_{i,j} - T_{i-1,j}  \right) \cdot
\lambda_{i-1/2,j} \cdot
\frac{r_{i-1/2}}{r_{i} - r_{i-1}} \cdot \frac{2}{r^{2}_{i+1/2} - r^{2}_{i-1/2}}
\nonumber \\
+
\left( T_{i,j+1} - T_{i,j}  \right) \cdot
\lambda_{i,j+1/2} \cdot
\frac{1}{\left( z_{j+1} - z_{j} \right) \cdot \left( z_{j+1/2} - z_{j-1/2}  \right)}
\nonumber \\
-
\left( T_{i,j} - T_{i,j-1}  \right) \cdot
\lambda_{i,j-1/2} \cdot
\frac{1}{\left( z_{j} - z_{j-1} \right) \cdot \left( z_{j+1/2} - z_{j-1/2}  \right)}.
\label{eq_int_main_3}
\end{eqnarray}
We define (for the simplest way)

$$
r_{i+1/2} = \frac{r_{i} + r_{i+1}}{2}, \qquad
r_{i-1/2} = \frac{r_{i-1} + r_{i}}{2},
$$

$$
z_{j+1/2} = \frac{z_{j} + z_{j+1}}{2}, \qquad
z_{j-1/2} = \frac{z_{j-1} + z_{j}}{2}.
$$
Thus for the rectangular grid Eq.~(\ref{eq_int_main_3}) can be written in the common form as

\begin{eqnarray}
\fl \rho_{i,j} C_{i,j} \frac{\hat{T}_{i,j} - T_{i,j}}{\tau}
\nonumber \\
= \frac{r_{i+1} + r_{i}}{r_{i+1} - r_{i}} \cdot
\frac{4 \lambda_{i+1/2,j}}{\left( r_{i+1} + r_{i} \right)^{2} - \lb r_{i} - r_{i-1} \rb^{2}}
\cdot \left[ T_{i+1,j} - T_{i,j}  \right]
\nonumber \\
-
\frac{r_{i} + r_{i-1}}{r_{i} - r_{i-1}} \cdot
\frac{4 \lambda_{i-1/2,j}}{\left( r_{i+1} + r_{i} \right)^{2} - \lb r_{i} - r_{i-1} \rb^{2}}
\cdot \left[ T_{i,j} - T_{i-1,j}  \right]
\nonumber \\
+
\frac{2 \lambda_{i,j+1/2}}{\lb z_{j+1} - z_{j} \rb \cdot \lb z_{j+1} - z_{j-1} \rb}
\cdot \left[ T_{i,j+1} - T_{i,j}  \right]
\nonumber \\
-
\frac{2 \lambda_{i,j-1/2}}{\lb z_{j} - z_{j-1} \rb \cdot \lb z_{j+1} - z_{j-1} \rb}
\cdot \left[ T_{i,j} - T_{i,j-1}  \right].
\end{eqnarray}
For constant space steps ($h_{r}$ and $h_{z}$) we have

\begin{eqnarray}
\fl \rho_{i,j} C_{i,j} \frac{\hat{T}_{i,j} - T_{i,j}}{\tau} =
\nonumber \\
\frac{\lambda_{i+1/2,j}}{h^{2}_{r}} \cdot \left( 1 + \frac{1}{2i}  \right) \left[ T_{i+1,j} - T_{i,j}  \right] -
\frac{\lambda_{i-1/2,m}}{h^{2}_{r}} \cdot \left( 1 - \frac{1}{2i}  \right) \left[ T_{i,j} - T_{i-1,j}  \right] +
\nonumber \\
+
\frac{\lambda_{i,j+1/2}}{h^{2}_{z}} \cdot \left[ T_{i,j+1} - T_{i,j}  \right] -
\frac{\lambda_{i,j-1/2}}{h^{2}_{z}} \cdot \left[ T_{i,j} - T_{i,j-1}  \right].
\label{main_eq}
\end{eqnarray}
To evaluate heat conductivity at the semi - integer domian points we use approximation \cite{patankar}

\begin{eqnarray}
\lambda_{i+1/2,j} = \frac{2 \lambda_{i,j} \lambda_{i+1, j}}{\lambda_{i,j} + \lambda_{i+1,j}},
\qquad
\lambda_{i-1/2,j} = \frac{2 \lambda_{i-1,j} \lambda_{i, j}}{\lambda_{i-1,j} + \lambda_{i,j}},
\nonumber \\
\lambda_{i,j+1/2} = \frac{2 \lambda_{i,j} \lambda_{i,j+1}}{\lambda_{i,j} + \lambda_{i,j+1}},
\qquad
\lambda_{i,j-1/2} = \frac{2 \lambda_{i,j-1} \lambda_{i,j}}{\lambda_{i,j} + \lambda_{i,j-1}}.
\label{lambda-aprox}
\end{eqnarray}
Equation (\ref{main_eq}) joins the temperature $T_{i,j}$ with $T_{i-1,j}$, $T_{i+1,j}$, $T_{i,j-1}$
and $T_{i,j+1}$, than Eq. (\ref{main_eq}) has 5-point space scheme and explicit scheme in time.
We will use 2-cycle scheme \cite{marchuk}. Let we rewrite Eq. (\ref{main_eq}) as follows

\begin{equation}
\rho_{i,j} C_{i,j} \frac{\hat{T}_{i,j} - T_{i,j}}{\tau} =
\Lambda_{r}T_{i,j} + \Lambda_{z}T_{i,j},
\end{equation}
where operators $\Lambda_{r,z}$ are defined by Eq. (\ref{main_eq}). The main idea of the 2-cycle
factorization scheme is making of the following steps ($k$ is time step index) \cite{marchuk}

\begin{equation}
\rho_{i,j} C_{i,j} \frac{T^{k-1/2}_{i,j} - T^{k-1}_{i,j}}{\tau} =
\Lambda_{r} \frac{T^{k-1/2}_{i,j} + T^{k-1}_{i,j}}{2}
\label{4_1}
\end{equation}

\begin{equation}
\rho_{i,j} C_{i,j} \frac{T^{k}_{i,j} - T^{k-1/2}_{i,j}}{\tau} =
\Lambda_{z} \frac{T^{k}_{i,j} + T^{k-1/2}_{i,j}}{2}
\label{4_2}
\end{equation}

\begin{equation}
\rho_{i,j} C_{i,j} \frac{T^{k+1/2}_{i,j} - T^{k}_{i,j}}{\tau} =
\Lambda_{z} \frac{T^{k+1/2}_{i,j} + T^{k}_{i,j}}{2}
\label{4_3}
\end{equation}

\begin{equation}
\rho_{i,j} C_{i,j} \frac{T^{k+1}_{i,j} - T^{k+1/2}_{i,j}}{\tau} =
\Lambda_{r} \frac{T^{k+1}_{i,j} + T^{k+1/2}_{i,j}}{2}
\label{4_4}
\end{equation}
The cycle of evaluation is namely to solve equations (\ref{4_1})-(\ref{4_4}). Let we rewite these
equations in a more useful form

\begin{eqnarray}
\fl \left[ B_{i,j}  \right] T^{k-1/2}_{i-1,j} -
\left[ A_{i,j} + B_{i,j} + 2\rho_{i,j} C_{i,j}/\tau  \right] T^{k-1/2}_{i,j} +
\left[ A_{i,j}  \right] T^{k-1/2}_{i+1,j}
\nonumber\\
= - \left(
\left[ B_{i,j}  \right] T^{k-1}_{i-1,j} -
\left[ A_{i,j} + B_{i,j} - 2\rho_{i,j} C_{i,j}/\tau  \right] T^{k-1}_{i,j} +
\left[ A_{i,j}  \right] T^{k-1}_{i+1,j}
\right)
\label{e1}
\end{eqnarray}

\begin{eqnarray}
\fl \left[ \Delta_{i,j}  \right] T^{k}_{i,j-1} -
\left[ \Delta_{i,j} + \Gamma_{i,j} + 2\rho_{i,j} C_{i,j}/\tau  \right] T^{k}_{i,j} +
\left[ \Gamma_{i,j}  \right] T^{k}_{i,j+1}
\nonumber\\
\lo = - \left(  
\left[ \Delta_{i,j}  \right] T^{k-1/2}_{i,j-1} -
\left[ \Delta_{i,j} + \Gamma_{i,j} - 2\rho_{i,j} C_{i,j}/\tau  \right] T^{k-1/2}_{i,j} +
\left[ \Gamma_{i,j}  \right] T^{k-1/2}_{i,j+1}
\right)
\label{e2}
\end{eqnarray}

\begin{eqnarray}
\fl \left[ B_{i,j}  \right] T^{k+1/2}_{i-1,j} -
\left[ A_{i,j} + B_{i,j} + 2\rho_{i,j} C_{i,j}/\tau  \right] T^{k+1/2}_{i,j} +
\left[ A_{i,j}  \right] T^{k+1/2}_{i+1,j}
\nonumber\\
\lo = - \left(  
\left[ B_{i,j}  \right] T^{k}_{i-1,j} -
\left[ A_{i,j} + B_{i,j} - 2\rho_{i,j} C_{i,j}/\tau  \right] T^{k}_{i,j} +
\left[ A_{i,j}  \right] T^{k}_{i+1,j}
\right)
\label{e3}
\end{eqnarray}

\begin{eqnarray}
\fl \left[ \Delta_{i,j}  \right] T^{k+1}_{i,j-1} -
\left[ \Delta_{i,j} + \Gamma_{i,j} + 2\rho_{i,j} C_{i,j}/\tau  \right] T^{k+1}_{i,j} +
\left[ \Gamma_{i,j}  \right] T^{k+1}_{i,j+1}
\nonumber\\
\lo = - \left(  
\left[ \Delta_{i,j}  \right] T^{k+1/2}_{i,j-1} -
\left[ \Delta_{i,j} + \Gamma_{i,j} - 2\rho_{i,j} C_{i,j}/\tau  \right] T^{k+1/2}_{i,j} +
\left[ \Gamma_{i,j}  \right] T^{k+1/2}_{i,j+1}
\right)
\label{e4}
\end{eqnarray}
where we denoted

\begin{equation}
B_{i,j} = \frac{\lambda_{i-1/2,j}}{h^{2}_{r}} \left( 1 - \frac{1}{2i}  \right),
\qquad
A_{i,j} = \frac{\lambda_{i+1/2,j}}{h^{2}_{r}} \left( 1 + \frac{1}{2i}  \right).
\end{equation}

\begin{equation}
\Delta_{i,j} = \frac{\lambda_{i,j-1/2}}{h^{2}_{z}},
\qquad
\Gamma_{i,j} = \frac{\lambda_{i,j+1/2}}{h^{2}_{z}}.
\end{equation}
The solving of these equations is strightforward, because equation matrixes are three-diagonal,
then these equations have 3-point space scheme. It is a very important note for our method,
because we will have possibility to decrease CPU time.

It might to prove 2-cycle method is absolutly time step stability \cite{marchuk} and it has
second-order in $\tau^{2}$ like Crank-Nicholson scheme and second-order in space.
Moreover defferential operators $\Lambda_{r,z}$ in the 2-cycle method can be noncommutating ones.

%%%%%%%%%%%%%%%%%%%%%%%%%%%%%%%%%%%%%%%%%%%
\section{Discretization of boundary conditions.}

\begin{figure}[h]
\begin{center}
\includegraphics[width=7.0cm]{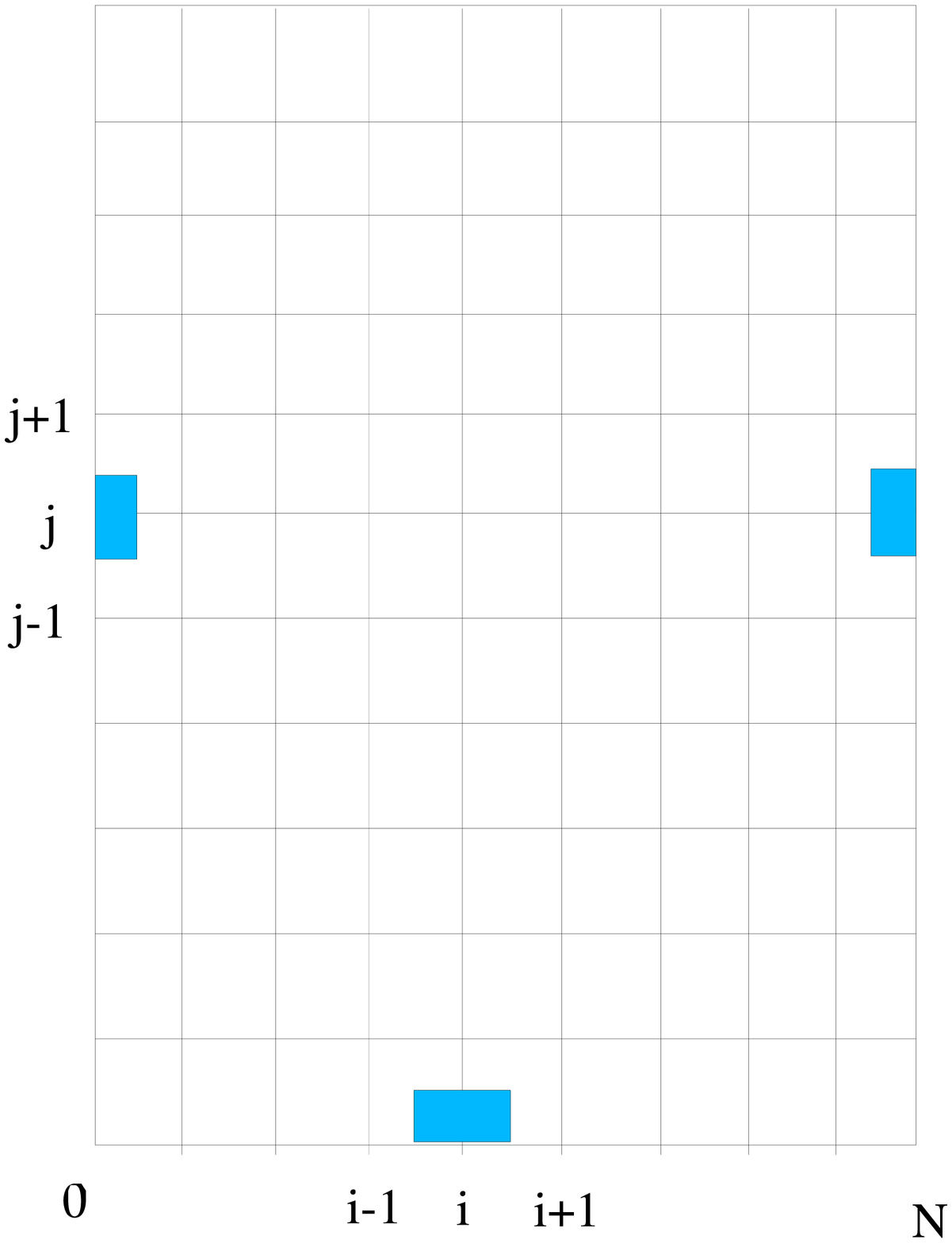}
\end{center}
\caption{\label{boundaries} The volumes for evaluating of integral balances in the boundary
conditions.}
\end{figure}

To solve the equations (\ref{e1})-(\ref{e4}) it is necessary to define values of temperature
on border or to set some relationships connecting temperature on border of a body with
temperature of an environment. Boundary conditions in a continuous limit are certain by equations
(\ref{cont-bc-AB}) - (\ref{cont-bc-FA}). We shall construct their discrete finite difference
analogues on the uniform grid, using a method of finite volume \cite{patankar}. For this purpose
we shall integrate the equation (\ref{main_eq_flux_PQ}) on some volume $G$,
adjoining to border of an ingot and on time within the limits of $[t, t + \tau]$. Such volumes
are shown on Fig.~\ref{boundaries}.

Integral balance of heat can be written as

\begin{equation}
\int\limits_{t}^{t+\tau} \int\limits_{G} \rho(T) C(T) \frac{\pd T}{\pd t} dV dt =
\int\limits_{t}^{t+\tau} \int\limits_{G}
\lb \frac{1}{r} \frac{\pd}{\pd r} (r Q) + \frac{\pd P}{\pd z} \rb
dV dt,
\label{int-bc-common}
\end{equation}
where heat fluxes are defined by Eqs. (\ref{flux_Q}) and (\ref{flux_P}).

After integration we shall receive the relationship connecting a thermal flux on border of a body,
set by a corresponding boundary condition, and thermal fluxes in nearby points for which some
approximations are required. Integration on time is led as follows: the left part of the
Eq.~(\ref {int-bc-common}) we approximate as follows 

\begin{equation}
\int\limits_{t}^{t+\tau} \int\limits_{G} \rho(T) C(T) \frac{\pd T}{\pd t} dV dt \sim
\hat{T} - T,
\end{equation}
and the right part of Eq.~(\ref{int-bc-common}) we shall calculate in {\em previous} the
moment of time.
% It is obvious, that at such {\it explicit} approximation of strongly nonlinear
% problems it is necessary to choose enough small time step $\tau$.
Now it is easy to calculate finite difference counterparts of boundary conditions.

\subsection{Boundary $AB$.}

Let area $G = [0, h _ {r}/2] \times [(j-1/2) h _ {z}, (j+1/2) h _ {z}] $ then the general expression
for a boundary condition on a site $AB $ after integration can be written down in the form of 

\begin{eqnarray}
\fl \frac{h_{r}^{2} h_{z}}{8}\rho C_{0,j} \left[ \hat{T}_{0,j} - T_{0,j} \right]
= \frac{\tau h_{z}}{2} \lambda_{1/2,j} \left[ T_{1,j} - T_{0,j} \right]
\nonumber\\
\lo + \frac{\tau h_{r}^{2}}{8 h_{z}} \lb \lambda_{0,j+1/2} \left[ T_{0,j+1} - T_{0,j} \right] -
\lambda_{0,j-1/2} \left[ T_{0,j} - T_{0,j-1} \right] \rb.
\label{dis-bc-AB}
\end{eqnarray}

\subsection{Boundary $BC$.}

On a site $BC $ there is no necessity to use a method of finite volume as in this case the field
of temperatures is set directly, therefore

\begin{equation}
\hat{T}_{i,M} = T_{L} + \Delta T.
\label{dis-bc-BC}
\end{equation}

\subsection{Boundary $CD$.}

On a site $CD$ we set the temperature, as well as on $BC$
\begin{equation}
\hat{T}_{i,M} = T_{L} + \frac{D_{cr} - 2ih_{r}}{D_{cr} - D_{el}} \Delta T.
\label{dis-bc-CD}
\end{equation}

\subsection{Boundary $DE$.}

Let $G = [(N-1) h _ {r}, Nh _ {r}] \times [(j-1/2) h _ {z}, (j+1/2) h _ {z}]$ then the general
expression for a boundary condition on a site $DE$ after integration it is possible to
write down in the form of 

\begin{eqnarray}
\fl \frac{h_{r}^{2} h_{z}}{8} (4N-1) \rho C_{N,j} \left[ \hat{T}_{N,j} - T_{N,j} \right] =
-\tau h_{z} N h_{r} q -
\tau h_{z} (N-\frac{1}{2}) \lambda_{N-1/2,j} \left[ T_{N,j} - T_{N-1,j} \right]
\nonumber \\
\lo + \frac{\tau h_{r}^{2}}{8 h_{z}} (4N-1)
\lb \lambda_{N,j+1/2} \left[ T_{N,j+1} - T_{N,j} \right] -
\lambda_{N,j-1/2} \left[ T_{N,j} - T_{N,j-1} \right] \rb.
\label{dis-bc-DE}
\end{eqnarray}

\subsection{Boundary $EF$.}

Let $G = [(N-1) h_{r}, N h_{r}] \times [(j-1/2) h_{z}, (j+1/2) h_{z}] $ then the general
expression for a boundary condition on a site $EF$ after integration it is possible to write
down area in the form of

\begin{eqnarray}
\fl \frac{h_{r}^{2} h_{z}}{8} (4N-1) \rho C_{N,j} \left[ \hat{T}_{N,j} - T_{N,j} \right]
\nonumber \\
\lo = - \tau h_{z} N h_{r} \beta^{cr}_{N,j} \left[ T_{N,j} - T_{cr} \right] +
\tau h_{z} (N-\frac{1}{2}) \lambda_{N-1/2,j} \left[ T_{N,j} - T_{N-1,j} \right]\
\nonumber \\
\lo + \frac{\tau h_{r}^{2}}{8 h_{z}} (4N-1)
\lb \lambda_{N,j+1/2} \left[ T_{N,j+1} - T_{N,j} \right] -
\lambda_{N,j-1/2} \left[ T_{N,j} - T_{N,j-1} \right] \rb
\label{dis-bc-EF}
\end{eqnarray}

\subsection{Boundary $AF$.}

Let $G = [(i-1/2)h_{r},(i+1/2)h_{r}] \times [0,h_{z}/2]$ then the general expression for a boundary
condition on a site $FA$ after integration it is possible to write down area in the form of 

\begin{eqnarray}
\fl \frac{i h_{r}^{2} h_{z}}{2} \rho C_{i,0} \left[ \hat{T}_{i,0} - T_{i,0} \right]
\nonumber \\
\lo = \frac{i \tau h_{r}^{2}}{h_{z}}
\lb
\lambda_{i,1/2} \left[ T_{i,1} - T_{i,0} \right] -
h_{z} \beta^{РПД}_{i,0} \left[ T_{i,0} - T_{РПД} \right]
\rb
\nonumber \\
\lo = \tau h_{z}
\lb
\lb i + \frac{1}{2} \rb \lambda_{i+1/2,0} \left[ T_{i+1,0} - T_{i,0} \right] -
\lb i - \frac{1}{2} \rb \lambda_{i-1/2,0} \left[ T_{i,0} - T_{i-1,0} \right]
\rb
\label{dis-bc-AF}
\end{eqnarray}
Further we use approximations for heat conductivity factors in semi-integer points, similar
to Eq.~(\ref {lambda-aprox})

%%%%%%%%%%%%%%%%%%%%%%%%%%%%%%%%%%%%%%%%%%
\section{The model parameters for VT3-1 alloy.}

As it is possible to see from expressions for boundary conditions on a mirror of a liquid bath
($BC$  and $CD$) we set distribution of temperature which is defined by alloy liquidus temperature
and an overheat depending on parameters of VAR. The overheat of the melt $\Delta$ can be
described by formula \cite{plavka}:

\begin{equation}
\Delta T(J, D_{in}) = 400 e^{-12\frac{D_{in}}{J}},
\end{equation}
where $J$ is arc current, kA; $D_{in}$ -- ingot diameter, m.
In a zone $DE$ the thermal flux $q$ to which measurements for alloy VT3-1 work
\cite{tet-64} is devoted is set. The parameter $\eta$ defines the relative contribution of each of
mechanisms of heat removal: at $\eta = 1$ the heat-conducting path goes only by radiation, and at
$\eta = 0$ only by convection. For $EF$ we accepted $\eta_{EF} = 1.0$, and for
$FA$ -- $\eta_{FA} = 0.5$. 

$C (T)$, including rate of release of latent heat in an interval of temperatures between liquidus
$T _ {L}$ and solidus $T _ {S}$, it is possible to present an effective specific thermal capacity
in the form of \cite{malevich}: 

\begin{equation}
\label{C}
\fl C(T) = \cases{
C_{L}(T) & for $T > T_{L}$ \\
g(T)C_{S}(T) + [1 - g(T)]C_{L}(T) - L\case{dg(T)}{dT} & for $T_{S}\leq T \leq T_{L}$\\
C_{S}(T) & for $T < T_{S}$\\}
\end{equation}
where $C_{L}$ and $C_{S}$ are specific heat capacities for liquid and solid phases; $L$ is
latent heat of fusion, $g(T)$ is solid fraction. In case of binary alloy for $g(T)$ we can obtain
simple expression (under lever rule suggestion)

\begin{equation}
g(T) = \frac{T_{m} - T_{S}}{T_{L} - T_{S}} \cdot \frac{T_{L} - T}{T_{m} - T}.
\end{equation}
For temperature derivative of $g(T)$ we can obtain

\begin{equation}
\frac{dg(T)}{dT} =
- \frac{T_{m} - T_{L}}{T_{L} - T_{S}} \cdot \frac{T_{m} - T_{S}}{(T_{m} - T)^2},
\end{equation}
where $T_{m}$ is fusion temperature of a solvent, $T_{L}$ is liquidus temperature, $T_{S}$
is solidus temperature. Parameters of model for alloy VT3-1 have been chosen by the following
$T_{m} = 1668 \; ^{o}C$, $T_{S} =  1550 \; ^{o}C$, $T_{L} = 1620 \; ^{o}C$, $L = 355000 \; J/kg$,
$T_{cr} = 70 \; ^{o}C$, $T_{bot} = 70 \; ^{o}C$, $\alpha_{bot} = 300 \; W/m^{2} \, K$.
Some data can be found in \cite{ti-sprav}. Position of a point $E$ was defined as in work
\cite{jardy-1}. 

For the account of increase in heat conductivity ТБУРМБЧБ due to fluid flow we
have increased heat conductivity of an alloy in a liquid phase \cite{jardy-1}.
Temperature dependence of the resulted ingot surface emissivity degree and
an internal surface of crucible has been chosen in the form of square-law dependence on
temperature \cite{ti-sprav}.

%%%%%%%%%%%%%%%%%%%%%%%%%%%%%%%%%%%%%%%%%%%%%%%%%%%%%%%%%%%%%%%%%%%%%%%%%%%%%%%%%%%%
\section{Simulation results and radiographical experiments.}

Using the mathematical model described above, we have led solidification modelling of ingots from
the alloy VT3-1. Modes of remeltings are specified in the Table \ref{table}. 

\begin{table}[b]
\caption{\label{table} Remelting regimes.}
\begin{indented}
\item[] \begin{tabular}{@{}ll}
\br
Ingot diameter, mm & Arc current, kA \\
\mr
750                & 37 \\
570                & 25 \\
435                & 15 \\
\br
\end{tabular}
\end{indented}
\end{table}

Some works \cite{plavka,radiography} describe the experimental research of the metal pool depth and
profile through fixation with radioactive isotopes. Below we compare the simulation results with
the experimental data on liquid metal pool depth at different points of time. In order to adequately
describe the experimental data we have carried out ''adjusting'' of the model. The adjusting provided for
rather exact match of the calculated and experimental metal pool depth, as well as liquid metal pool
profiles at different ingot height. Due to the fact that the most extensive experimental data are
available for 750 mm dia ingot melted at 37 kA, we carried out ''adjusting'' for the above-mentioned
melting mode. The main variable parameters were -- heat conductivity in the liquid phase
(that was considered not dependent on temperature) and the surface emissivity factor depending
on temperature.

\subsection{\o 750 mm ingot. Arc current $J = 37 \; kA$.}

\begin{figure}[h]
\begin{center}
\includegraphics[width=9.0cm]{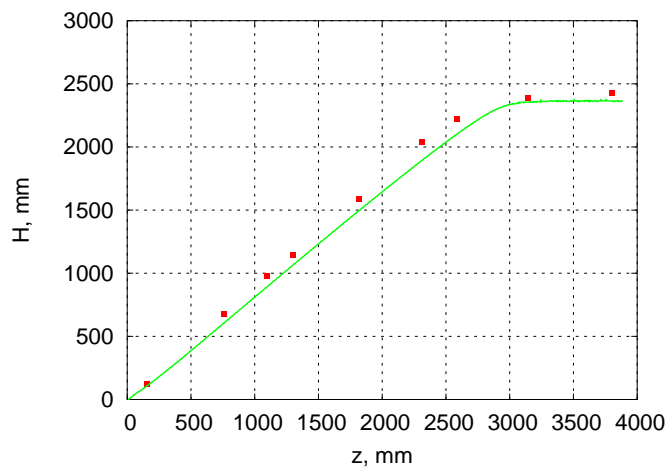}
\end{center}
\caption{\label{750-37-comparison} Experimental and theoretical pool depth versus the height
of the melted ingot. \o 750 mm
ingot, arc current $J = 37 \; kA$. Points -- experiment, curve -- theory.}
\end{figure}

One of the most important parameters assessed during analysis of one or another VAR mode for
titanium alloys (as well as for nickel-base, iron-base and zirconium base alloys
\cite{jardy-1,jardy-2,bertram,malevich}) is the depth of the liquid metal pool.
Figure~\ref{750-37-comparison} shows the theoretical metal pool depth with experimentally
obtained depth (through fixation with tungsten radioactive isotopes) depending on the height of
the melted ingot. As is obviously, the theory describes the metal pool depth behavior satisfactorily.
During melting the quasi-steady state was achieved at the metal pool depth having been steady.

The description of the liquid metal pool only {\it may be misleading} as the pool depth does not
actually reflect the volume of the liquid metal at the particular melt point.  In order to assess
feasibility of satisfactory simulation of the liquid metal pool volume using the model, we
calculated the liquid metal pool profiles at the points of time recorded on the experimental
radiogram. This shows that the model adequately simulates the profiles of the liquid metal
pool during the whole melt process. All this together suggests that on the basis of the simulation
approach in question it is feasible to simulate some most important parameters of metal
solidification under VAR (width of mushy zone, temperature gradient, isotherm travel rate).

\begin{figure}[ht]
\begin{center}
\includegraphics[height=10.0cm]{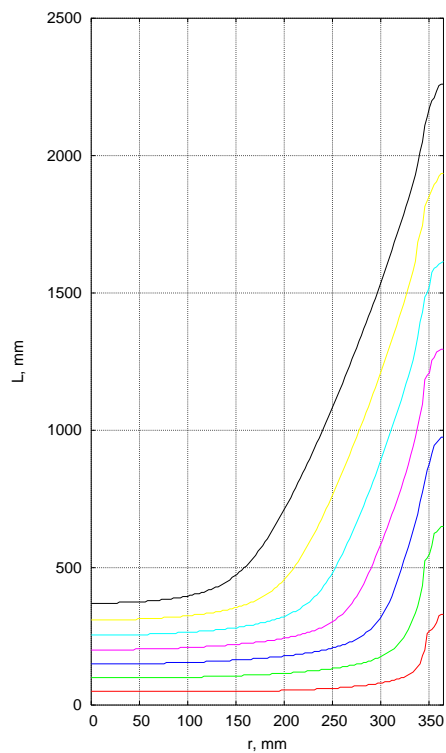}
\end{center}
\caption{\label{profiles-20-minute} Calculated profiles of the liquid metal pool during first
140 minutes of melting (with 20 minute interval).}
\end{figure}

Figure~\ref{profiles-20-minute} shows the calculated profiles of the liquid metal pool during
first 140 minutes of melting (with 20 minute interval) to evaluate the linear solid-melt
interface travel rate in different ingot zones. The comparison of the simulated profiles with
the experimentally obtained metal pool profiles shows that the simulated
metal pool volumes do not deviate from the experimental ones by more than 15\%.

\subsection{\o 570 mm ingot. Arc current $J = 25 \; kA$.}

\begin{figure}[t]
\begin{center}
\includegraphics[width=9.0cm]{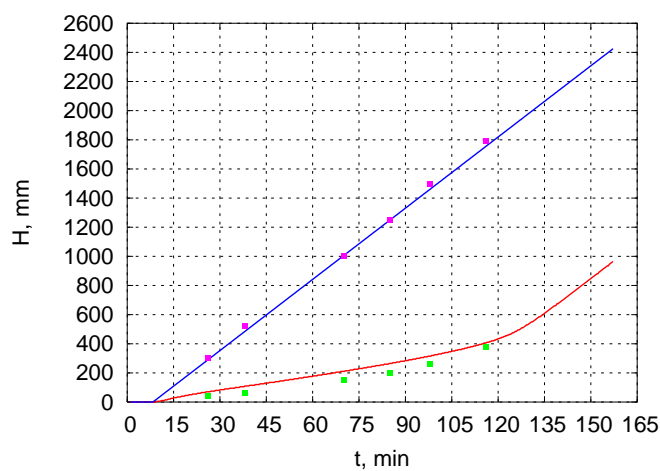}
\end{center}
\caption{\label{570-25-comparison} Location of the pool open surface (upper curve and points) and
pool bottom coordinate versus the melting time. \o 570 mm ingot, arc current $J = 25 \; kA$.
Theory -- curves, experiment -- points.}
\end{figure}

Figure~\ref{570-25-comparison} shows the theoretical curves and experimental points for
$570\;mm$ ingot melted at 25 kA. In this case we compared the simulated data with the
experimental data according to location of the pool open surface (i.e. melt mirror) and
pool bottom coordinate versus the melting time. It can be seen from Figure~\ref{570-25-comparison}
that the quasi-steady state was not achieved.

\subsection{\o 435 mm ingot. Arc current $J = 15 \; kA$.}

\begin{figure}[b]
\begin{center}
\includegraphics[width=9.0cm]{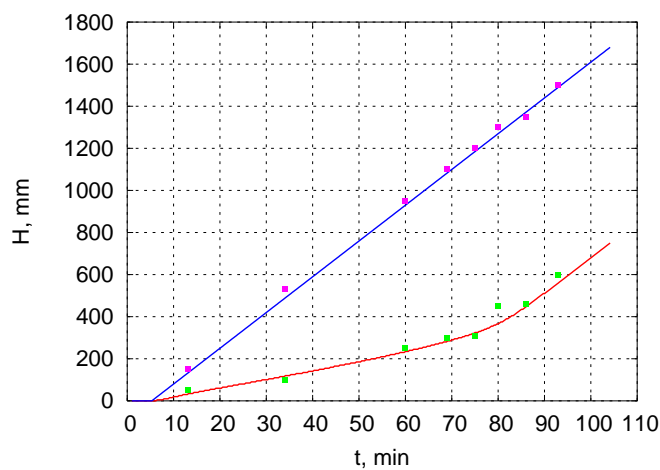}
\end{center}
\caption{\label{435-15-comparison} Location of the pool open surface
(upper curve and points) and pool bottom coordinate versus the melting time. \o 435 mm
ingot, arc current $J = 15 \; kA$.
Theory -- curves, experiment -- points.}
\end{figure}

Figure~\ref{435-15-comparison} shows the theoretical curves and experimental points for
435 mm ingot melted at 15 kA. In this case as above we compared the simulated data with the
experimental data according to location of the pool open surface (i.e. melt mirror) and
pool bottom coordinate versus the melting time. It can be seen from
Figure~\ref{435-15-comparison} that the quasi-steady state was achieved.

Therefore, a conclusion can be made that under {\it selected} parameters for Vt3-1 alloy,
the simulation model can describe the main behavior tendencies of such VAR parameters as
liquid metal pool depth and profile at different melting points. It is very important to note
that the model that was ''adjusted'' once, gives satisfactory description of the
experimental data beyond the ''adjusting values''.

%%%%%%%%%%%%%%%%%%%%
\section{Conclusion.}

This article describes the mathematical simulation model of heat processes that take place under
VAR. The discretization has been made of the {\em non-linear} heat conductivity equation through
the Marchuck's method of integral identities and of {\em non-linear} boundary conditions through
the final volume method. The solution algorithm in question has unconditional numerical
stability and is applicable for tasks with non-commuting differentiation operators dependent
on time. 

For the purposes of the simulation model testing we calculated the liquid metal pool depth
under VAR melting of titanium alloy VT3-1. It was observed that the model simulates the
liquid metal pool (depth and profiles) rather adequately during the whole melting process for
different ingot diameters and different current strength. The relative error in determining
the liquid metal pool profile does not exceed 15\%. This would enable to use the model in
future to calculate various melting modes in order to assess such parameters as: liquid metal
pool depth, liquid metal pool volume, width of two-phase zone, temperature gradient,
isotherms travel rate and local solidification time.

\end{document}